\documentstyle[12pt]{article}
\begin{document}
\input amssym.def
\input amssym

\font\twelverm=cmr12
\font\sixrm=cmr6        
\font\eightrm=cmr8
\font\caps=cmbx8
\font\eightsl=cmsl8

\font\gross=cmcsc10 scaled\magstep 1    
\font\gr=cmcsc10      
\font\symb=cmbxti10    
\font\bi=cmmib10

\font\twelvebf=cmbx12    
\font\twbf=cmb10 scaled \magstep 1

\font\tit=cmti10 scaled\magstep 2      
\font\Titel=cmr10 scaled \magstep 2

\hyphenation{
Meas-ure-ment Meas-uring 
meas-ure-ment meas-ure meas-uring
pre-meas-ure-ment pre-meas-ure pre-meas-uring}

\def\sy{system}                       \def\py{physic}
\def\qu{quantum}                      \def\mec{mechanic}
\def\qm{quantum mechanics}
\def\ty{theory}                       \def\qt{\qu\ \ty}
\def\me{meas\-ure}                    \def\mt{\me{}ment} \def\mts{\mt{}s}

\def\pov{{\gross pov} meas\-ure}      \def\pv{{\gross pv} \me}
\def\pvo{sharp \ob}           \def\povo{unsharp \ob}
\def\stv{{\gross stv} meas\-ure}     \def\stf{state transformer}
\def\prob{probability}

\def\op{operator}                    \def\ops{operators}
\def\ob{observable}                  \def\obs{observables}
\def\sad{self-adjoint}               \def\sop{\sad\ \op}
\def\Rea{{\bold R}}                     \def\eps{\varepsilon}
\def\Re{{\bold R}}
\def\fii{\varphi}  \def\vart{\vartheta} \def\ck{$\cal K$}
\def\hi{{\cal H}}                     \def\ph{\cal P(\cal H)}
\def\of{(\Omega ,\cal F)}       \def\rb{\bigl(\Rea ,\cal B(\Rea )\bigr)}
\def\pa{\cal R_{\cal S}}         \def\pas{\cal R_{\cal A}} 
\def\s{$\cal S$}         \def\a{$\cal A$}      \def\sa{$\cal S + \cal A$}

\def\ist{insolubility theorem}
\def\os{(\Omega,\Sigma)}

\def\wt{\widetilde{\rho}_{\cal A}}


\title{Can `Unsharp Objectification' Solve
the Quantum Measurement Problem?\thanks{
Proceedings of the Quantum Structures 1996 Meeting,
Berlin, 29.7.--3.8.1996.
\newline\hbox{\hskip 15pt} {\sl Int. J. Theor. Phys.}
{\bf 36}, 1997. (Submitted November 1996).}}
 
\author{Paul Busch\\
{\rm Department of Mathematics,
The University of Hull, UK}\\
{\rm e-mail: p.busch@maths.hull.ac.uk}
}
\date{30 November 1996}
\maketitle

\begin{abstract}
The quantum measurement problem is formulated in the form of
an insolubility theorem that states the impossibility of
obtaining, for all available object preparations, 
a mixture of states of the compound object and
apparatus system that would represent definite pointer
positions. A proof is given that comprises arbitrary object
observables, whether sharp or unsharp, and besides sharp
pointer observables a certain class of unsharp pointers,
namely, those allowing for the property of
{\sl pointer value definiteness}.
A recent result of H. Stein is applied to allow for the
possibility that a given measurement may not be applicable
to all possible object states but only to a subset of them.
The question is raised whether the statement of the
insolubility theorem remains true for genuinely unsharp
observables. This gives rise to a precise notion of {\sl
unsharp objectification}. 
\end{abstract}

\section*{\centerline{\twbf 1. Introduction}}

 The claim of the insolubility of the quantum measurement problem has
been given a precise formulation in a series of papers
aiming at increasing generality of the premisses (see, eg.,
Wigner, 1963, d'Espagnat, 1966, Fine, 1970, Shimony, 1974, Brown, 1986).
The most recent step provided an extension of the
insolubility proof to include measurements of arbitrary
sharp or unsharp object observables (Busch and Shimony, 1996).
In the present
contribution I consider the even more general case of measurements based on
pointer observables that are not necessarily sharp. It will
be shown that the established proof
strategy of the previous no-go theorems can be adapted so as
to cover a certain class of unsharp pointer observables:
those admitting definite values. Technically this
corresponds to the case of a positive operator valued ({\gr
pov}) measure $Z$ on a $\sigma$-algebra $\Sigma$ which is
such that each effect $Z(X)$, $X\in\Sigma$, has eigenvalue
1. The corresponding eigenstates are those states in which
the pointer has {\sl definite values}. 
Note that this does not require the effect to be a
projection. 
It will be shown that no unitary measurement exists
in which the compound object plus apparatus system 
could always (i.e., for arbitrary initial object states) 
be in a mixture of states in which the extended pointer
($I\otimes Z$)  would have definite values. The occurrence
of such a mixture is a necessary condition for the
{\sl pointer objectification\/} (Busch, Lahti and Mittelstaedt,
1996). The proof technique
used here differs from that used by Shimony (1974) and Busch
and Shimony  (1996) in
that a recent theorem due to H. Stein (1997) is
applied. This provides an extension of his impossibility
theorem and the insolubility theorem Busch and Shimony (1996). 

This result implies that the quantum measurement problem is
not simply due to idealisations in which possible
measurement inaccuracies are neglected: in using the general
representation of observables as {\gr pov} measures, all  kinds
of inaccuracy have been taken into account -- to the extent  they 
are still compatible with the idea of definite pointer values.
The remaining potential loophole is furnished by the case of
pointer observables which are {\sl genuinely unsharp} in
that they do not allow for pointer value definiteness. 
This opens up the challenge to make precise sense of the
idea of {\sl unsharp objectification} which will be done 
here.

Apart from the possibility that no insolubility theorem
might hold for genuinely unsharp pointers, the existing
no-go theorems allow an exhaustive systematic overview of
the possible modifications of quantum mechanics, or of its
interpretations, that may be, and have been, undertaken to
resolve (or dissolve) the measurement problem
(Busch, Lahti and Mittelstaedt, 1996).

\section*{\centerline{\twbf 2. Notion of Measurement in
Quantum Mechanics}}

In the following I shall adopt the usual Hilbert space
formulation of quantum mechanics where observables and
states  are
represented as, and identified with, certain positive
operator valued ({\gross pov}) measures and density
operators, respectively. These concepts are required to 
formulate the probability structure of the theory in its
(probably) most general form. Then according to
the minimal interpretation of quantum mechanics, the
probability measures provided by the formalism give the
probability distributions for measurement outcomes,
and thus the expected experimental statistics.

This minimal notion of an observable -- and of a measurement -- 
is captured in the so-called {\sl probability
reproducibility} condition. The essential elements of a
measurement are conveniently summarised in the concept of a
{\sl measurement scheme}, represented as a quadruple
${\cal M}:=\langle{\cal H}_{\cal A}, \rho_{\cal A}, U,Z\rangle$,
where ${\cal H}_{\cal A}$ denotes the Hilbert space of the 
measuring device (or probe) ${\cal A}$, $Z$ the pointer
observable of $\cal A$,
i.e., a {\gross pov} measure on some measurable space
$(\Omega,\Sigma)$, $\rho_{\cal A}$ a fixed initial state of
$\cal A$, and $U$ the unitary measurement coupling serving
to establish a correlation between the object system $\cal
S$ (with Hilbert space $\cal H$) and $\cal A$. Any
measurement scheme $\cal M$ fixes a unique observable of
$\cal S$, that is, a {\gross pov} measure $E$ on $(\Omega,\Sigma)$
such that the following condition is fulfilled:
\begin{itemize}
\item {\bf Probability Reproducibility Condition:}
$$
{\rm tr}[I\otimes Z(X)\,U\rho\otimes\rho_{\cal A}U^*]
\ =\ {\rm tr}[E(X)\rho]\eqno{\rm (PR)}
$$
for all states $\rho$ of $\cal S$ and all outcome sets
$X\in\Sigma$. 
\end{itemize}
$E$ is the observable measured by means of 
$\cal M$. Conversely, if an observable $E$ of $\cal S$ is
given, then this condition determines which measurement
schemes $\cal M$  serve as measurements of $E$.

\section*{\centerline{\twbf 3. The Objectification Problem}}

The probability reproducibility condition specifies what it
means that a measurement scheme serves to measure a certain
observable. However, this condition does not exhaust the
notion of measurement. In fact the reproduction of
probabilities in the pointer statistics requires first of
all that in each run of a measurement a pointer reading will
occur; in other words: it is part of the notion of
measurement that measurements do have definite outcomes.
While the concept of a measurement scheme allows one to {\sl
describe} what happens to the object and apparatus when an
outcome arises, quantum meachanics is facing severe
difficulties to {\sl explain} the occurrence of such
outcomes. This problem arises if one starts with the
interpretational idea that an observable has a definite
value when the object system in question is in an eigenstate
of that observable. If a probe system is coupled to that
object, then probability reproducibility requires that the
corresponding value is indicated with certainty by the
pointer reading after the measurement interaction has
ceased. In this way a definite value of the object
observable leads deterministically to a definite value of
the pointer observable. However, if the object is {\sl not}
in an eigenstate, the observable cannot be ascertained to
have a definite value, and by the linearity of the unitary
measurement coupling, the compound object plus probe system
ends up in a state in which it cannot be ascertained, by
appeal to the eigenstate-eigenvalue link, that
the pointer has a definite value. This is the measurement
problem, or the problem of the objectification of pointer
values. 

Resolutions to this problem are being sought by changing the
rules of the game: either on the side of the formalism
(introduction of classical observables, or modified
dynamics), or on the interpretational side (hidden variables
theories such as `Bohmian mechanics', or various `no-collapse'
interpretations). Before embarking on such radical
revisional programmes, it seems fair to make sure that the 
measurement problem is not merely a consequence of overly
idealised assumptions that would disappear in a more realistic
account. It turns out, however, that the problem does
persist even when measurements are allowed to be inaccurate
and the measuring system is in a mixed rather than a pure
state. The development of these arguments is reviewed in 
Busch and Shimony (1996), where an insolubility theorem is given that
pertains to measurements of sharp and unsharp object
observables. This result has recently been overtaken 
by H. Stein (1997) who showed that the objectification problem
persists for arbitrary measurement schemes also when the
measurement is not required to be applicable to {\sl all}
object preparations but only to states in some subspace of the
object's Hilbert space. Based on this result, a
further  step will now be
taken that comprises the possibility of the pointer
being an unsharp observable as well, 
as long as pointers can still assume definite values.

In order to give the precise statement of the insolubility
theorem, let us consider a measurement scheme $\cal M$.  The
theorem is based on the following requirements as 
necessary conditions for the definiteness, or objectivity,  
of sharp values of the pointer $Z$ in the postmeasurement state
$$
\rho'_{\cal SA}\equiv U\,\rho_{\cal S}\otimes\rho_{\cal
A}\,U^*.
$$ 

\begin{itemize}
\item{\bf Pointer mixture condition}:
$$
\rho'_{\cal SA}\ =\sum I\otimes Z(x_i)^{1/2}\rho'_{\cal SA}
\,I\otimes Z(X_i)^{1/2}\ \equiv\ \sum \rho'_{\cal SA}(X_i)
\eqno{\rm (PM)}
$$
for some partition $\Omega=\cup X_i$ and all initial object
states $\rho$;
\item {\bf Pointer value definiteness}:
$$
{\rm tr}\big[I\otimes Z(X_i)\,\rho'_{\cal SA}(X_i)\big]
\ =\ {\rm tr}\big[\rho'_{\cal SA}(X_i)\big] 
\eqno{\rm (PVD)}
$$
for all $i$ and all initial object
states $\rho$.
\end{itemize}
For a derivation of these conditions, see (Busch, Lahti and
Mittelstaedt, 1996).
The first says that the postmeasurement state should be a
mixture of pointer eigenstates, while the second requires
that the final states conditional on reading a result in
$X_i$ are indeed eigenstates of the pointer for which $X_i$
has probability one to occur again upon immediate repetition
of the reading of the pointer observable $Z$. 

\vskip 6pt
\noindent
{\bf Insolubility Theorem.} 
If a measurement scheme 
$\cal M$ fulfills (PM) and (PVD) for all object states $\rho$
supported in some subspace $\hi_0$ of $\hi$, 
then the measured observable $E$
according to (PR) is {\sl trivial} with respect to all such states; that is,
${\rm tr}\big[E(X)\,\rho\big]
=\lambda(X)$ for all $X\in\Sigma$, where $\lambda$ is
a state-independent probability measure on
$(\Omega,\Sigma)$. Hence if a  measurement scheme is to lead
to objective pointer values, it will yields no
information at all about the object.

\section*{\centerline{\twbf 4. Proof of the Insolubility Theorem}}

We make use of the following lemma by H. Stein (1997),
applying it very much in the same way as Stein himself did
but using our terminology and allowing for unsharp pointers.

\noindent
{\bf Lemma.} Let $Q,R$ be bounded linear operators on $\cal
H_{\cal A}$ and $\cal H\otimes\cal H_{\cal A}$,
respectively. Let $\cal H_0$ be a vector subspace of $\cal
H$. Assume that for all nonzero vectors $\fii\in\cal H_0$,
$$
\big(P[\fii]\otimes Q\big)\,R\ =\ R\,\big(P[\fii]\otimes
Q\big).
$$ 
(Here $P[\fii]$ denotes the projection onto the ray
containing $\fii$.) Then there exists a unique bounded
linear operator $\wt$ in $\cal H_{\cal A}$ such that 
$$
\big(P[\fii]\otimes Q\big)\,R\ =\ P[\fii]\otimes\wt\qquad
{\rm for\ all\ }\fii\in\hi_0.
$$

We apply this as follows: for any $\fii\in\hi_0$ we denote
$\rho_{\cal SA}(\fii):=
P[\fii]\otimes\rho_{\cal A}$, and 
$\rho_{\cal S\cal A}'(\fii):=
U\,\big(P[\fii]\otimes\rho_{\cal A}\big)\,U^{-1}$.
By assumption (PVD), any nonzero effect $Z(X)$ has
eigenvalue 1.
Let $Z(X_i)^{(1)}$ denote the corresponding spectral projection of the
effect $Z(X_i)$. Then the assumption (PM) is equivalent to
saying that each nonzero component state $\rho_{\cal SA}'(X_i)$ is
an eigenstate of $Z(X_i)^{(1)}$ associated with the
eigenvalue 1, that is, $Z(X_i)^{(1)}\rho_{\cal SA}'(X_i)=
\rho_{\cal SA}'(X_i)$, for all $X_i$ of the given partition.
Therefore (PM) implies that
$\rho_{\cal S\cal A}'(\fii)$ commutes with all
$I\otimes Z(X_i)^{(1)}$, and thus also with all $Z(X_i)$:
$$
\left[I\otimes Z(X_i)\, ,\,\rho_{\cal SA}'(\fii)\right]\ =\ O.
$$
 We rewrite this as follows:
$$
\left[U^{-1}\big(I\otimes Z(X_i)\big)\,U\, ,\,
P[\fii]\otimes\rho_{\cal A}
\right]\ =\ {O}.
$$

Now we make the following choices for the operators $R,Q$ 
introduced in the Lemma: for each $i$, let
$R_i=U^{-1}\big(I\otimes Z(X_i)\big)\,U$ and $Q_i=
\rho_{\cal A}$. Then by virtue of the Lemma there exists 
an operator $\wt(X_i)$ such that
$$
P[\fii]\otimes\rho_{\cal A}\,U^{-1}\big(I\otimes
Z(X_i)\big)\,U\ =\ P[\fii]\otimes \wt(X_i).
$$
Taking the trace yields the probabilities for the measured
observable $E^{\cal M}$:
$$
{\rm tr}\big[P[\fii]\,E^{\cal M}(X_i)\big]\ =\ 
{\rm tr}\big[\wt(X_i)\big].
$$
As the operators $\wt(X_i)$ are independent of $\fii$, it
follows that the measured observable is trivial with respect
to states from the subspace $\hi_0$. This completes the proof.

\section*{\centerline{\twbf 5. Unsharp Objectification}}

The residual question left open by the above result is
whether the conclusion of `no information gain' remains
valid if the assumption (PVD) of definite pointer values is
dropped. That is, one would only require a modified form of
(PM) to hold: the final object-plus-apparatus state should
be a mixture of states,
$$
\rho_{\cal SA}'\ =\ \sum_i\rho_{\cal SA}''(X_i),
$$
 in which the pointer is {\sl unsharply real}. By this we
mean that the component states should be `near-eigenstates'
of $I\otimes Z(X_i)$ in the sense that they give
probabilities close to one for the corresponding $X_i$. If
in addition in can be ascertained that the above mixture
admits an ignorance interpretation, then it shall be said
that {\sl unsharp objectification} has taken place.

Unsharp objectification, as explained here, would be a rather natural
option if the pointer observables available in realistic experiments were
genuinely unsharp (so that they would not allow for probabilities
equal to one). One can argue that pointers, being macroscopic
quantities, are in fact of that kind. Some of the arguments
supporting this conclusion are detailed in (Busch, Lahti and
Mittelstaedt, 1996) and (Busch, Grabowski and Lahti, 1995).
Unsharp pointer readings correspond to a situation where the pointer
states associated with different values are not (strictly) orthogonal.
Thus one cannot claim with certainty that the reading one means to
have taken is reproducible on a `second look' at the pointer. For
macroscopic quantities, however, the potential error will be practically
negligible as it can be extremely small compared to the scale of the
reading.

Yet I would
conjecture that unsharp objectification cannot be achieved
either. Once this would have been established, one could
safely conclude that the only way out lies in some of the
mentioned modifications either of the formalism or the
interpretation of quantum mechanics. Nevertheless it seems worthwhile
to pursue the notion of unsharp pointers as it may contribute to
resolving some problems these alternative approaches are still facing,
such as the so-called tail problem that arises in the case of the
(continuous) spontaneous collapse models.

\section*{\centerline{\twbf Acknowledgement}}

For reasons which those who were present at my talk
will understand I felt ``forced'' by the Chairman's presence
to finish my talk five minutes early. 
(``A {\sl watched speaker} never ends late.'')
I am grateful for this
to have happened (the responsibility for which is all mine) 
as it happily induced me to elaborate
(at least to my satisfaction) those parts of my manuscript 
that I ``had to'' skip in my talk. 

\section*{\twbf References}

Brown, H.~R. (1986). {\sl Found.\ Phys.} {\bf 16}, 857.

 \vskip 3pt
\noindent 
Busch, P., Grabowski, M., and  Lahti, P. (1995).
{\sl Operational Quantum Physics}, Springer-Verlag,
Berlin.

\vskip 3pt
\noindent
Busch, P., Lahti, P.~J., and Mittelstaedt, P. (1996).
{\sl The Quantum Theory of 
Measurement}, Springer-Verlag, Berlin, Second Revised Edition.

\vskip 3pt
\noindent
Busch, P., and Shimony, A. (1996). Insolubility of the Quantum
Measurement Problem for Unsharp Observables, 
{\sl Stud.\ Hist.\ Mod.\
Phys.} {\bf 27}, in press.

\vskip 3pt
\noindent
d'Espagnat, B. (1966). {\sl Nuovo Cim.\ Suppl.} {\bf 4}, 828.

\vskip 3pt
\noindent
Fine, A. (1970). {\sl Phys.~Rev.~D} {\bf 2}, 2783.

\vskip 3pt
\noindent
Shimony, A. (1974). {\sl Phys.~Rev.~D} {\bf 9}, 2321.

\vskip 3pt
\noindent
Stein, H. (1997). On a maximal impossibility theorem
for the quantum theory of measurement. In: {\sl Nonlocality,
Passion at a Distance, and Entanglement: Quantum Mechanical
Studies for Abner Shimony}, eds. B.~Cohen, M.A.~Horne,
J.~Stachel, Elsevier, Dordrecht.

\vskip 3pt
\noindent
Wigner, E.~P. (1963).  {\sl Am.\ J.\ Phys.} {\bf 31}, 6.

\end{document}